\documentclass[11pt]{article}
\usepackage{hyperref}
\pdfoutput=1

\begin{document}
\title{Droplet break-up with negative momentum \\ Fluid Dynamics Videos}

\author{Laurent Tanguy, Dong Liang, Roland Zengerle and Peter Koltay \\
\vspace{6pt}\\
IMTEK - University of Freiburg,\\
Georges-Koehler-Allee 103, 79110 Freiburg, Germany\\
HSG-IMIT - Institut für Mikro- und Informationstechnik, \\
Georges-Koehler-Allee 103, 79110 Freiburg, Germany\\
BioFluidix GmbH, Georges-Koehler-Allee 103, 79110 Freiburg, Germany}

\maketitle


\section{Introduction}

The ejection of liquid droplets from a nozzle is highly important field of fluid dynamics. 
The Weber number describes the amount of kinetic energY that is needed to overcome the surface tension and tO create a free-flying droplet with positive momentum. 
According to literature Weber numbers above 12 [1] assure the creation and the safe break up of liquid droplets. 
However, even when Weber numbers decrease below 8, it is still possible to observe droplet break-up sometimes with particular effects.
We present here fluid dynamics videos for droplet break-up at low Weber numbers where the supplied energy is smaller than the surface energy of the generated droplet, resulting in a negative momentum of the droplet. 
This situation is there characterized by the droplet formation, break-up from the nozzle and then return of the droplet back into the nozzle. 
The inverted droplet momentum is due to the fact that during the droplet formation the surface tension begins to slow down the flow velocity inside the droplet and finally inverts the flow direction and the average momentzum of the droplet, while the droplet tail still breaks off from the nozzle due to necking an a local positiv flow at the tail of the droplet. 
Thus after the break-up finallky occures the average droplet momentum is already oriented toward the nozzle. 
It is therefore possible to observe the droplet returning into the nozzle.

\section{Materials and set-up}

In order to experimentally observe this rare phenomenon, we used a PipeJet P9 dispenser from BioFluidix, GmbH [2]. This dispenser is based on a piezo stack actuated piston providing a precisely controlled displacement to a 500 micrometer wide liquid filled polyimide tube with open end (nozzle).
The fluid inside the tube is pushed out of the nozzle by direct displacement, creating droplets with adjustable volume between 15 and 60 nl, provided the kinetic energy - controlled by the displacement velocity - is high enough to overcome surface tension. 
While the droplet velocity is controlled by the displacement velocity, the volume of the droplets can be adjusted by the magnitude of the displacement.
In order to observe droplet break-up with negative momentum, one starts with a regular droplet ejection at an arbitrary setting of displacement and velocity that leads to stable droplet production. Then the velocity of the piston is incrementally reduced by small steps. Thus, the kinetic energy transmitted to the droplet can be carefully controlled, while the droplet mass remains approximately constant. When the total supplied kintetik energy is dropping slightly below the surface energy of the droplet, still a droplet break-up can occure. However, in this case the total droplet momentum has to become negative to maintain energy conservation.\\
A high-speed camera was used to record the droplet formations, breaks-up and free flights under several experimental conditions. The framerate was 10602 fps for each experiment.
The results of the experiments can be seen in the linked videos where droplets of different volumes (i.e. different piston displacement) were created first with positive momentum, then with negative momentum where they can be seen returning back into the nozzel, and finally no droplet break-up when the piston velocity is still further reduced.
For each droplet volume the displacement velocity has to be precisely adjusted, as the phenomenon only occurs only for a narrow range of velocities.

\section{Video}

The video shows that drop-on-demand droplet break-up can also be achieved at Weber numbers below 8 (considering the nozzle diameter as characteristic dimension). This conditions is characterized by the fact that the kinetic energy alone is not sufficient to separate the droplet from the bulk fluid. 
The protruding liquid jet reaches a state where its kinetic energy falls to zero and the velocity of the jet head begins to revert towards the nozzle direction due to the surface tension slowing down the jet. If surface tension is high and the droplet tail thin, like in the presented experiments, the droplet tail can still neck-off to form a temporarily free flying droplet, while the average droplet momentum is already inverted.
Since the total momentum of the droplet is already inverted at that time, the droplet returns back into the nozzle and merges with the bulk fluid. This interpretation is supported by computational fluid dynamics simulations using the Volume of Fluid method to model surface tension effects. In 2D-simulations performed with decreasing flow velocities at the nozzle, droplet break-up with negativ momentum can be observed with the same characteristics like in the experiment. The velocity field vizualized inside the droplet clearly indicates that the momentum (i.e. flow direction) is already inverted at the time the droplet finally breaks off. As shown in the videos there is a good qualitative agreement between the simulation and the experiments.

This video has been submitted to the Gallery of Fluid Motion 2012 which is an annual showcase of fluid dynamics videos.

\section{References}
1 - M. Pilch and C.A. Erdman, \textit{Use of breakup time data and velocity history data to predict the maximum size of stable fragments for acceleration-induced breakup of a liquid drop}, International Journal of Multiphase Flow, vol. 13, 6, pp. 741-757, 1987.\\
2 - W. Streule, T. Lindemann, G. Birkle, R. Zengerle, P. Koltay, \textit{PipeJet: A Simple Disposable Dispenser for the Nano- and Microliter Range}, Journal of the Association for Laboratory Automation, vol. 9, pp. 300-306, 2004.

\end{document}